\documentclass[twocolumn,aps,prl,superscriptaddress]{revtex4-2}
\usepackage[utf8]{inputenc}
\usepackage{amsmath}
\usepackage{stix2}
\usepackage{bm}
\usepackage{xcolor}
\usepackage{graphicx}

\begin{document}

\title{Fractional Chern Insulator in Twisted Bilayer MoTe$_2$}

\author{Chong Wang}
\affiliation{Department of Materials Science and Engineering, University of Washington, Seattle, WA 98195, USA}
\author{Xiao-Wei Zhang}
\affiliation{Department of Materials Science and Engineering, University of Washington, Seattle, WA 98195, USA}
\author{Xiaoyu Liu}
\affiliation{Department of Materials Science and Engineering, University of Washington, Seattle, WA 98195, USA}
\author{Yuchi He}
\affiliation{Rudolf Peierls Centre for Theoretical Physics, Clarendon Laboratory, Parks Road, Oxford OX1 3PU, United Kingdom}
\author{Xiaodong Xu}
\affiliation{Department of Physics, University of Washington, Seattle, WA 98195, USA}
\affiliation{Department of Materials Science and Engineering, University of Washington, Seattle, WA 98195, USA}
\author{Ying Ran}
\affiliation{Department of Physics, Boston College, Chestnut Hill, MA, 02467, USA}
\author{Ting Cao}
\email{tingcao@uw.edu}
\affiliation{Department of Materials Science and Engineering, University of Washington, Seattle, WA 98195, USA}
\author{Di Xiao}
\email{dixiao@uw.edu}
\affiliation{Department of Materials Science and Engineering, University of Washington, Seattle, WA 98195, USA}
\affiliation{Department of Physics, University of Washington, Seattle, WA 98195, USA}

\begin{abstract}
A recent experiment has reported the first observation of a zero-field fractional Chern insulator (FCI) phase in twisted bilayer MoTe$_2$ moir\'e superlattices~\cite{cai_signatures_2023}. The experimental observation is at an unexpected large twist angle 3.7$^\circ$ and calls for a better understanding of the FCI in real materials.  In this work, we perform large-scale density functional theory calculation for the twisted bilayer MoTe$_2$, and find that lattice reconstruction is crucial for the appearance of an isolated flat Chern band. The existence of the FCI state at $\nu = -2/3$ is confirmed by exact diagonalization. We establish phase diagrams with respect to the twist angle and electron interaction, which reveal an optimal twist angle of $3.5^\circ$ for the observation of FCI. We further demonstrate that an external electric field can destroy the FCI state by changing band geometry and show evidence of the $\nu=-3/5$ FCI state in this system. Our research highlights the importance of accurate single particle band structure in the quest for strong correlated electronic states and provides insights into engineering fractional Chern insulator in moir\'e superlattices.
\end{abstract}

\maketitle

\textit{Introduction}.---Fractional Chern insulators (FCI), the analog of the fractional Quantum Hall effect~\cite{PhysRevLett.48.1559,PhysRevLett.50.1395} in lattice systems, feature fractional excitations with anyonic statistics~\cite{parameswaran2013fractional,bergholtz2013topological,neupert2015fractional,liu_recent_2023}. Due to its exotic nature and potential applications in topological quantum computing~\cite{PhysRevLett.94.166802,RevModPhys.80.1083}, the FCI has long been sought after in condensed matter experiments. To find FCI, the widely-accepted approach is to find a Chern band with significantly quenched kinetic energy~\cite{PhysRevLett.106.236804,PhysRevLett.106.236802,sheng2011fractional,PhysRevX.1.021014,xiao2011interface}. This can be realized in two-dimensional (2D) moir\'e superlattices, which have been shown to be a fruitful and tunable platform to explore electronic correlation~\cite{andrei2021marvels, kennes2021moire}. Indeed, theoretical proposals of FCI have been put forward in twisted bilayer graphene~\cite{PhysRevResearch.2.023237,PhysRevResearch.2.023238,PhysRevLett.124.106803,PhysRevB.103.125406,PhysRevLett.126.026801,PhysRevB.104.L121405} and twisted bilayer transition metal dichalcogenide (TMD)~\cite{li2021spontaneous,crepel2022anomalous,morales2023pressure}. Experimental evidences for FCI have also been found in graphene-based superlattices~\cite{xie2021fractional,spanton2018observation}, albeit at a finite magnetic field. Recently, the first observation of the FCI in the absence of magnetic field (i.e., the fractional quantum anomalous Hall effect) has been reported in twisted bilayer MoTe$_2$ at hole fillings $\nu = -2/3$ and $\nu = -3/5$~\cite{cai_signatures_2023}. The observation is at an unexpected large twist angle ($\sim 3.7^\circ$), for which the Chern bands had been commonly believed to be too dispersive to stabilize the FCI. The experimental observation calls for a better understanding of single-particle band structure of twisted TMD bilayer as well as the emergence of the FCI in real materials.

In this work, we perform large-scale density functional theory (DFT) calculation for the twisted bilayer MoTe$_2$. In contrast to previous theoretical studies, we find that the band structure features an isolated flat Chern band that favors FCI. By exact diagonalization (ED) calculations, we confirm the existence of FCI at $\nu= -2/3$. We also find that the ferromagnetism at $\nu = -1/3$ is much weaker than that at $\nu = -2/3$, explaining the absence of the $\nu = -1/3$ FCI in the experiment. We further investigate the fate of FCI under an external electric field and find that FCI become unstable at $E = 1.26$~mV/\AA{}, which is consistent with experimental observation. The suppression of FCI by external electric field is attributed to the deterioration of the flatness of band geometry. We establish phase diagrams with respect to the twist angle and electronic interaction, revealing an optimal twist angle of $3.5^\circ$ for the observation of FCI. Finally, we show evidences of the $\nu=-3/5$ FCI state in this system. Our research highlights the importance of accurate single particle band structure in the quest for strong correlated electronic states and provides insights into engineering FCI in moir\'e superlattices.

\textit{Single-particle electronic structure}.---
Moir\'e superlattices formed by twisting a bilayer introduce a long wavelength periodic structure characterized by the moir\'e lattice constant $a_M=a/\theta$, where $a$ is the lattice constant of the original 2D layer and $\theta$ is the twist angle. This large length scale makes it possible to model the low-energy electronic structure with a continuum model. The valence band edge of monolayer MoTe$_2$ is located at the corners of its Brillouin zone, i.e., $K$ and $K'$ points. The two points are separated by a large momentum such that $K$ and $K'$ valleys can be considered independently. Following the experiment~\cite{cai_signatures_2023}, we consider $R$ stacking twisted bilayer and the continuum model Hamiltonian for $K$ valley reads~\cite{wu_topological_2019,yu2020giant}
\begin{equation}
    \mathcal{H}_K = \left(
        \begin{array}{cc}
            H_b & \Delta_T(\bm{r}) \\
            \Delta_T^{\dagger}(\bm{r}) & H_t
        \end{array}
    \right).
\end{equation}
Here, $H_{b/t} = -\hbar^2\left(\bm{k}-\bm{K}_{b/t}\right)^2 / 2 m^* + \Delta_{b/t}(\bm{r})$ is the bottom ($b$) and top ($t$) layer Hamiltonian subjected to a moir\'e potential $\Delta_{b/t}(\bm{r})=2 v \sum_{j=1,3,5} \cos \left(\bm{G}_j \cdot \bm{r} \pm \psi\right),$ where the bottom (top) layer corresponds to the positive (negative) sign. $\bm{K}_{b/t}$ is the $K$ point for the bottom and top layer and $\bm{G}_j$ is the moir\'e reciprocal lattice vectors defined by $\bm{G}_j = \frac{4\pi}{\sqrt{3}a_M} (\cos(\frac{\pi (j-1)}{3}), \sin(\frac{\pi (j-1)}{3}))$. The interlayer tunneling is dictated by three-fold rotational symmetry as $\Delta_T(\boldsymbol{r}) = w\left(1+e^{-i \boldsymbol{G}_2 \cdot \boldsymbol{r}}+e^{-i \boldsymbol{G}_3 \cdot \boldsymbol{r}}\right)$. $m^* = 0.6 m_e$ is the effective mass ($m_e$ is the bare electron mass) and $(v, \psi, w)$ are the free parameters in the continuum model.  The continuum Hamiltonian for the $K'$ valley can be obtained by applying time reversal symmetry to $\mathcal{H}_K$. Inside each valley, the electrons are fully spin polarized due to the large spin-valley coupling, and opposite valleys have opposite spin due to time-reversal symmetry~\cite{PhysRevLett.108.196802,cao2012valley}.

\begin{table}
\begin{tabular}{|c|c|c|c|}
\hline
 &  $v$ (meV) & $\psi$ ($^\circ$) & $w$ (meV)  \\
 \hline
 Local-stacking approx.~\cite{wu_topological_2019} & 8.0 & -89.6 & -8.5 \\
 \hline
 Large-scale DFT & 20.8 & +107.7  & -23.8 \\
 \hline
\end{tabular}
\caption{Parameters for the continuum model.\label{table:parameters}}
\end{table}

\begin{figure}
\centering
\includegraphics[width=0.942\columnwidth]{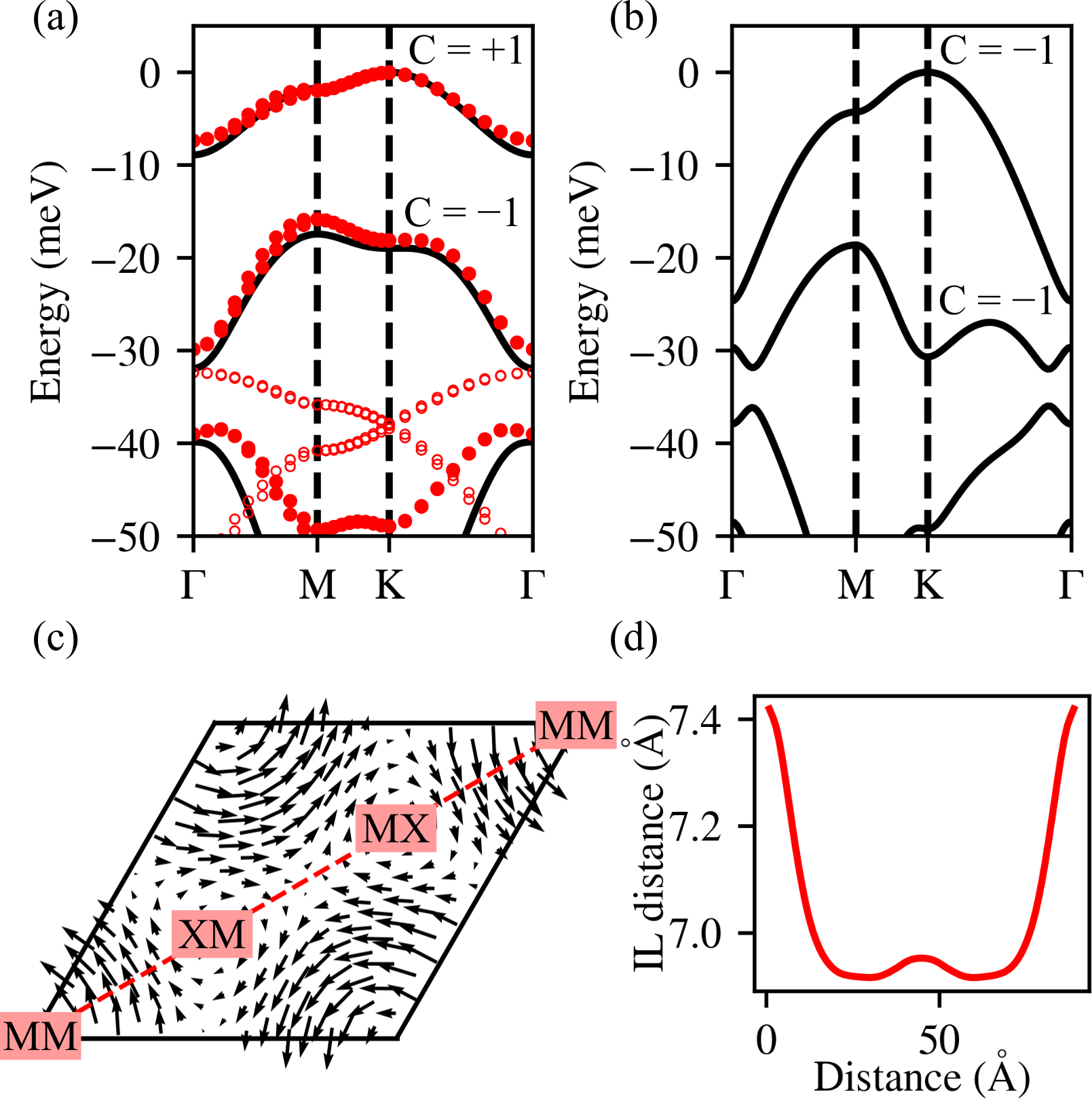}
\caption{Band structures for $K$ valley calculated by continuum model with parameters derived from our DFT calculation [(a)] and parameters from Ref.~\onlinecite{wu_topological_2019} [(b)]. Chern numbers for the two topmost bands are labeled in the plot. Kohn-Sham DFT band structure is plotted with red circles in (a), and the two DFT bands labeled by empty circles are from the $\Gamma$ valley. The twist angle is 3.89$^\circ$. (c) shows in-plane atomic displacement field in a moir\'e unit cell after relaxation, and (d) shows interlayer (IL) distance for the line cut in (c). The maximal in-plane atomic displacement is 0.32~\AA{}. High symmetry stackings are labeled in (c). MM/XM/MX denotes the stacking where the metal/chalcogen/metal atoms of the top layer are directly above the metal/metal/chalcogen atoms of the bottom layer, respectively. \label{fig:bands}}
\end{figure}

The parameters can be fixed by various approaches, among which the simplest one is to fit from DFT calculations for bilayer MoTe$_2$ at various stackings. This approach is adopted by Ref.~\onlinecite{wu_topological_2019} and the parameters are reproduced in Table~\ref{table:parameters}. The valence band structure with this set of parameters is shown in Fig.~\ref{fig:bands}(b) at twist angle $3.89^\circ$. The topmost valence band is dispersive with band width larger than 20~meV. In addition, the two topmost bands overlap each other in energy. Both features are unfavorable for the emergence of FCI.

In this work, we seek to establish a better understanding of the single particle band structure by performing large-scale DFT calculations to take into account atomic relaxation, layer corrugation and interlayer electric polarization (details in the Supplemental Material~\cite{supplemental}). We choose the closest commensurate twist angle (3.89$^\circ$) to the experimental value and construct the moir\'e superlattice of MoTe$_2$ using its monolayer unit cell with the optimized lattice constant $a=3.52$~\AA{}. The band structure of the moir\'e superlattice is presented in Fig.~\ref{fig:bands}(a) as red dots. The DFT result shows significant lattice reconstruction in both in-plane and out-of-plane direction [Fig.~\ref{fig:bands}(c-d)]. We then fit the continuum model parameters to the DFT band structure, and the result is presented in Table~\ref{table:parameters}. Compared to the parameters from Ref.~\onlinecite{wu_topological_2019}, our parameters features a much larger moir\'e potential and interlayer tunneling, which is likely caused by the significant lattice reconstruction~\cite{PhysRevLett.121.266401,li2021lattice,li2021imaging}, resulting in an isolated Chern band with band width of roughly 9~meV [cf. Fig.~\ref{fig:bands}(a)].

\begin{figure}
\centering
\includegraphics[width=0.949\columnwidth]{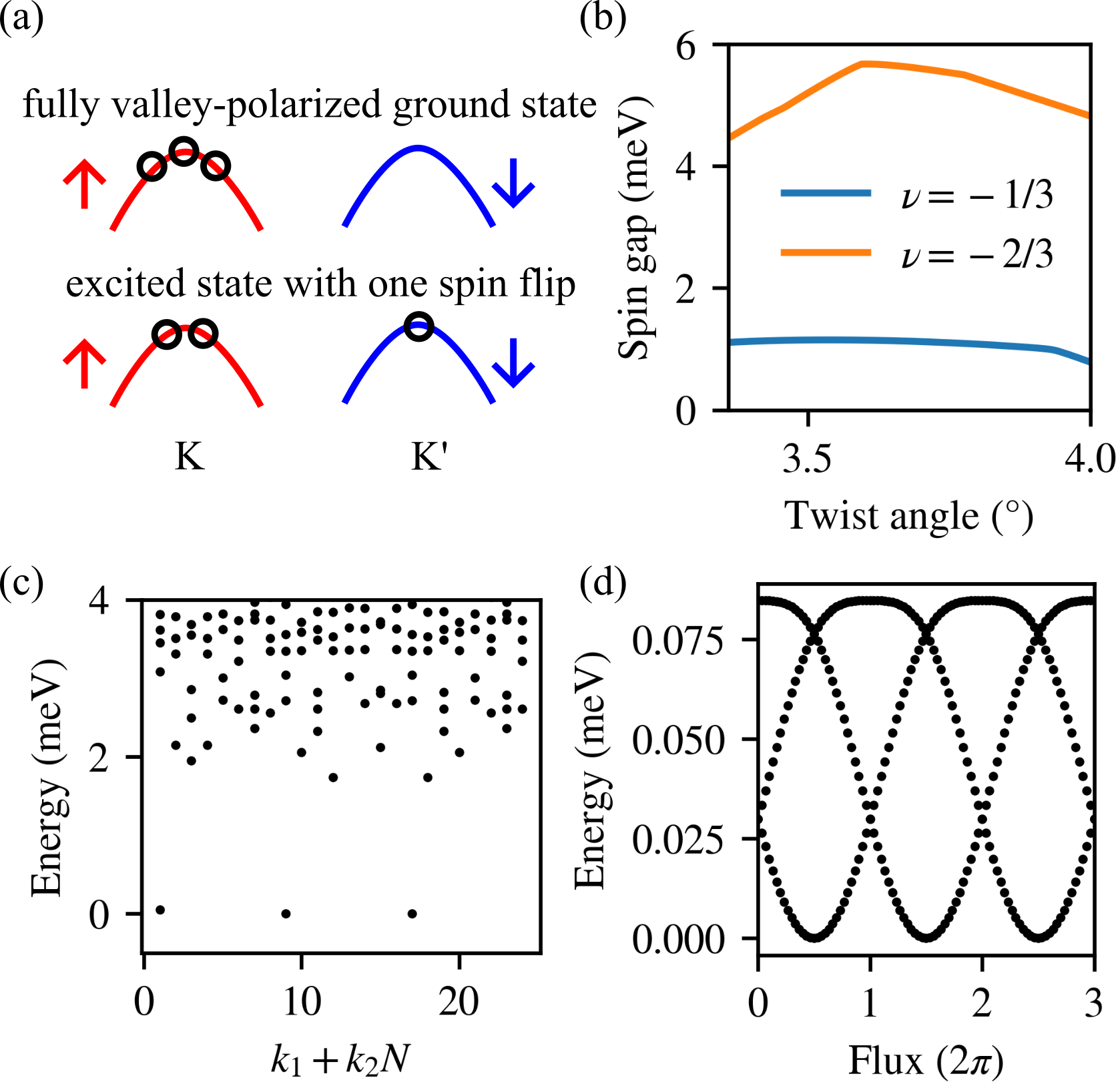}
\caption{For a large range of twist angles, the ground state of twisted bilayer MoTe$_2$ is fully valley polarized, as shown in the top panel of (a). The arrows represent spin, which is locked to the valley ($K$ and $K'$) degree of freedom. The bottom panel of (a) shows an example of excited state with one spin flip. The energy difference between the lowest state with spin flip(s) and the ground state is defined as the spin gap, shown in (b) for $\nu = -1/3$ and $\nu = -2/3$ as a function of the twist angle. The calculation for (b) is carried out with $3 \times 4$ unit cells. (c) The many-body spectrum as a function of total crystalline momentum with the assumption of full valley polarization at $\nu=-2/3$. (d) shows the evolution of ground states under flux insertion along the $k_2$ direction. During the flux insertion, the many-body gap is maintained. The calculation for (c-d) is carried out with $4 \times 6$ unit cells; the dielectric constant is chosen to be 15; the distance between gate and sample is chosen to be $d = 300$~\AA{}; the twist angle is 3.89$^\circ$.
\label{fig:fci}}
\end{figure}

\textit{Fractional Chern insulator at $\nu=-2/3$}.---Having established the existence of an isolated, relatively flat Chern band, we investigate whether FCI can be stabilized. We adopt the following form of the Coulomb interaction:
\begin{equation}
H_{\text{int}} = \frac{1}{2A} \sum_{l, l', \tau, \tau', \bm{k}, \bm{k}', \bm{q}} V (\bm{q}) {c_{l \tau \bm{k}+\bm{q}}^{\dagger}}  {c_{l' \tau' \bm{k}' -\bm{q}}^{\dagger}}  {c_{l' \tau' \bm{k}'}} c_{l \tau\bm{k}},
\end{equation}
where $V(\bm{q}) = e^2\text{tanh}(|q| d) / 2\epsilon_0 \epsilon |q|$ is the Coulomb interaction with dual-gate screening, $A$ is the area of the system (proportional to the number of $k$ points in the calculations), $d$ is the distance between the twisted bilayer MoTe$_2$ and two symmetric metal gates, $\epsilon_0$ is the vacuum permittivity and $\epsilon$ is the relative dielectric constant. Here, $c_{l \tau \bm{k}}^\dagger$ creates a plane wave with momentum $\bm{k}$ at valley $\tau$ and layer $l$. Due to spin-valley locking, $\tau$ can also be understood as the spin label.  We project the interaction onto the topmost moir\'e band and carry out ED calculations. We choose $\epsilon=15$ to make the characteristic interaction strength smaller than the averaged energy gap. For smaller $\epsilon$, we present the phase diagram in the Supplemental Material~\cite{supplemental}. While we left a more accurate treatment to include the band mixing for future studies, there are evidences that FCI can still be stablized even when the interaction exceeds the band gap~\cite{PhysRevLett.112.126806,PhysRevB.91.035136}.

The precursor to the FCI is spontaneous time-reversal symmetry breaking. We first perform ED calculations taking both valleys into account with a system size of $3 \times 4$ unit cells. We find that over a broad range of twist angles, the ground state for both $\nu = -1/3$ and $\nu = -2/3$ is fully valley-polarized, with holes occupying only one valley [cf. Fig.~\ref{fig:fci}(a)].  Since the spin and valley indices are coupled, full valley polarization implies full spin polarization. The spin gap, defined as the energy difference between the lowest-energy state that does not exhibit full valley-polarization and the fully valley-polarized ground state, is shown in Fig.~\ref{fig:fci}(b) for both $\nu = -1/3$ and $\nu = -2/3$. The spin gap for $\nu = -1/3$ is much smaller than that of $\nu = -2/3$, indicating much weaker ferromagnetism of the former. The difference in the spin gap is consistent with the experimental observation that ferromagnetism appears at $\sim$4.5 K at $\nu = -2/3$, whereas no ferromagnetism is observed at $\nu = -1/3$ down to base temperature of 1.6~K~\cite{communications}. %Therefore, we focus on $\nu = -2/3$ in the following.

Given the large spin gap and strong ferromagnetism at $\nu = -2/3$, we further carry out ED calculations for a single valley, which allows us to consider a larger system with $4 \times 6$ unit cells.

The most important signature of the FCI is the ground state degeneracy when the system is put on a torus~\cite{PhysRevB.31.3372,PhysRevB.41.9377}. Indeed, the ED energy spectrum shows three nearly degenerate states, separated by an energy gap from other states, as shown in Fig.~\ref{fig:fci}(c). Under flux insertion, the three ground states evolve into each other, exhibiting a $6\pi$ periodicity [Fig.~\ref{fig:fci}(d)]. We also calculate many-body Chern number~\cite{PhysRevB.31.3372} at this filling to be $-2/3$, which is consistent with the experimental observation~\cite{cai_signatures_2023}. The single particle occupation number, defined as $n(\bm{k}) = \langle c_{\bm{k}}^\dagger c_{\bm{k}} \rangle$ is presented in the Supplemental Material~\cite{supplemental}. The uniformity of $n(\bm{k})$ is a strong indicator favoring FCI over charge density wave (CDW) states, one of FCI's competing phases. The above evidences provide strong evidence to the existence of FCI in twisted MoTe$_2$ at $\nu = -2/3$.

\begin{figure}
\centering
\includegraphics[width=0.879\columnwidth]{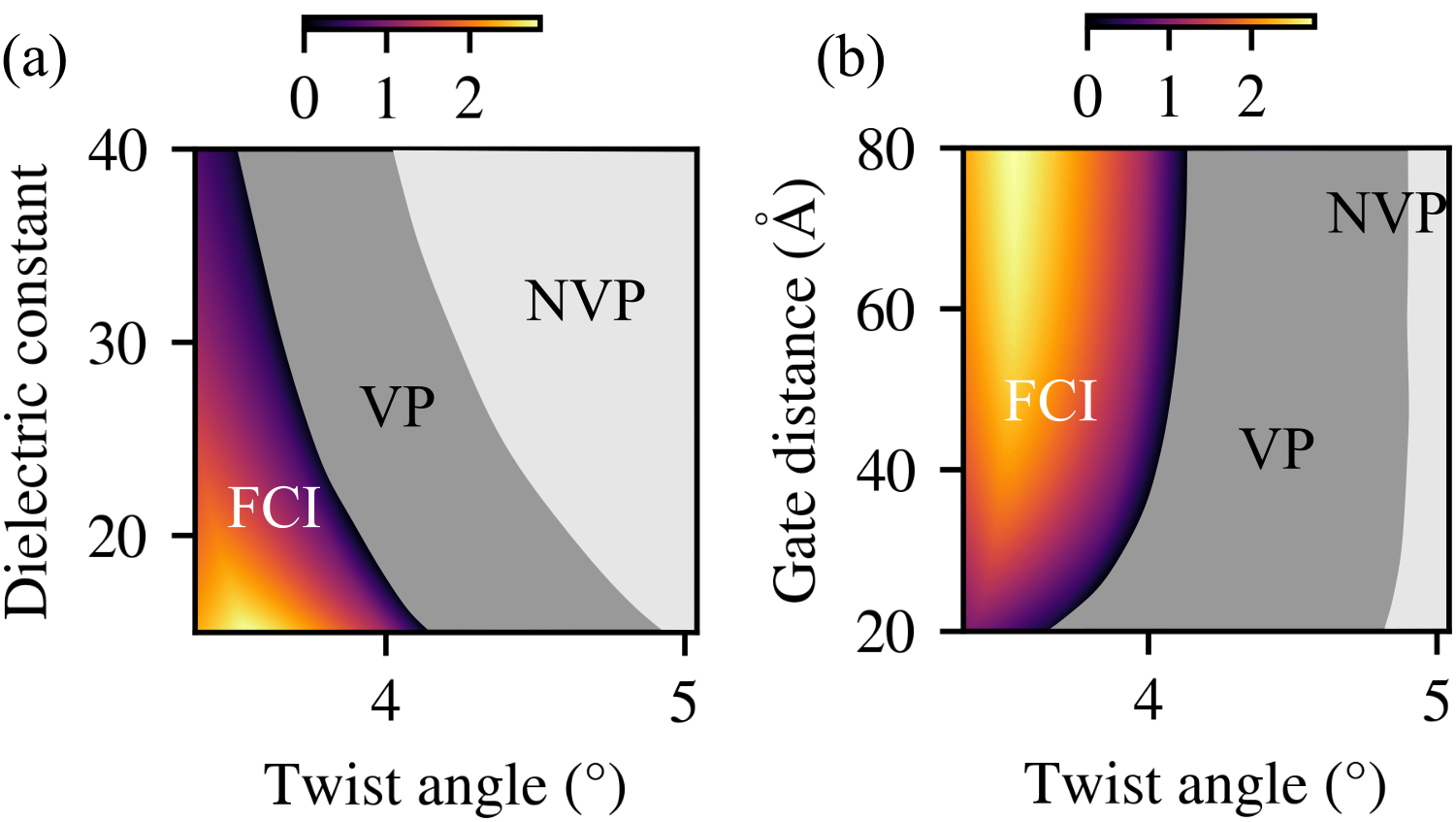}
\caption{(a) Phase diagram as a function of twist angle and dielectric constant at $\nu=-2/3$. $d$ is chosen as $300$~\AA{}. FCI: fractional chern insulator; VP: valley-polarized state; NVP: non-valley-polarized state. (b) Phase diagram as a function of twist angle and $d$ at $\nu=-2/3$. $\epsilon$ is chosen as 15. Many-body gap is shown in both (a) and (b) by color in FCI phase (unit: meV). FCI is identified with $4 \times 6$ unit cells. Valley polarization is identified with $3 \times 4$ unit cells.
\label{fig:phase_diagram}}
\end{figure}

\textit{Phase diagram}.---The emergence of the FCI state depends on the dominance of electron-electron interaction energy over single particle kinetic energy. This ratio between the two energy scales can be adjusted by two factors: the single-particle band width, and the  environmental dielectric screening of electron-electron interactions. The most experimentally accessible knob to tune the band width is changing the twist angle. For example, in twisted bilayer graphene, flat band emerges at a series of magic angles~\cite{bistritzer2011moire}. For twisted bilayer TMD systems, the band width is less sensitive to twist angles and we find relative isolated flat bands for twist angle $3^\circ-4^\circ$. The screening of the electron-electron interactions can be tuned by changing $\epsilon$, as well as  the distance between the sample and the metal gate $d$: larger $d$ leads to weaker screening of the electron-electron interactions.

The phase diagram, as a function of $\theta$ and either $\epsilon$ or $d$, is presented in Fig.~\ref{fig:phase_diagram}. Within the range of the twist angle presented in Fig.~\ref{fig:phase_diagram}, the band width of the topmost valence band increases monotonically with $\theta$. There are three distinct phases: the FCI phase, the valley-polarized (VP) phase, and the non-valley-polarized (NVP) phase. The NVP phase emerges in the weak electron-electron interaction regime, characterized by large $\theta$ and large $\epsilon$ [see Fig.\ref{fig:phase_diagram}(a)], or small $d$ [see Fig.\ref{fig:phase_diagram}(b)]. For stronger interactions, all holes occupy the same valley, which is shown as the VP phase in Fig.\ref{fig:phase_diagram}. Our numerical evidence suggests that the VP states are most likely metal states with fully polarized spin, meaning they represent a half-metal state~\cite{crepel2022anomalous}. The nature of these VP phases is left for further studies. FCI emerges from the VP phases at even stronger interactions. The many-body gap for FCI phase has a peak at $\theta=3.5^\circ$, indicating an optimal twist angle for the observation of FCI. This optimal angle is close to the twist angle ($\sim 3.7^\circ$) of the device in which FCI is observed~\cite{cai_signatures_2023}.

\begin{figure}
\centering
\includegraphics[width=0.970\columnwidth]{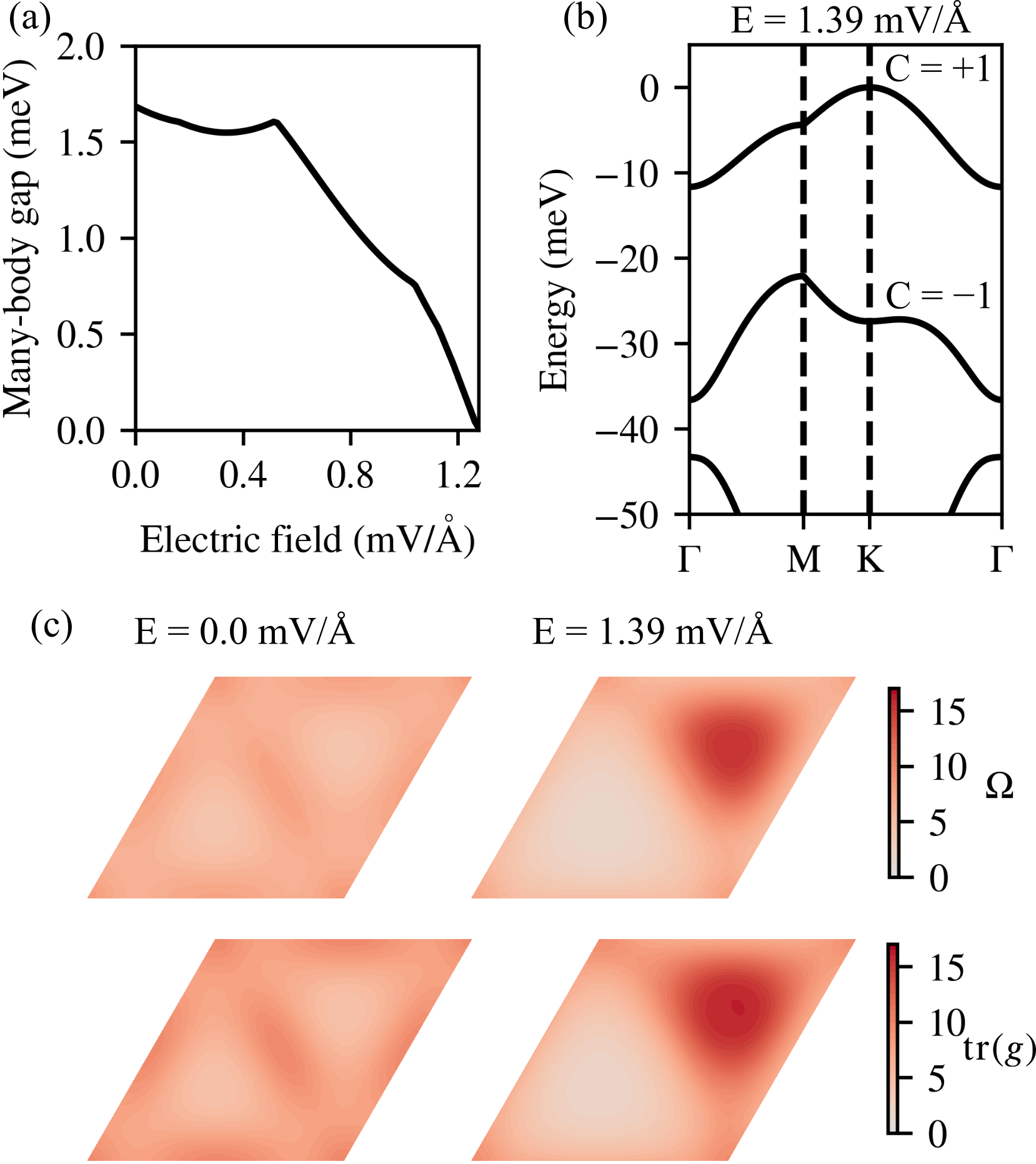}
\caption{(a) Many-body gap as a function of external electric field at $\epsilon = 15$ and $\theta = 3.89^\circ$. (b) Band structure at $E = 1.39$~mV/\AA{} and $\theta = 3.89^\circ$ (FCI is already destroyed by the electric field). (c) Indicators for the emergence of FCI at $E = 0$~mV/\AA{} (left) and $E = 1.39$~mV/\AA{} (right). $\Omega$ is the Berry curvature and $\mathrm{tr}(g)$ is the trace of quantum metric tensor. The unit for (c) is the inverse of the area of the moir\'e Brillouin zone.
\label{fig:electric_field}
}
\end{figure}

\textit{The effect of electric field}.---An out-of-plane electric field generates potential differences between the top and bottom layers. Experimentally, it is observed that the FCI at $\nu = -2/3$ can be suppressed by this out-of-plane electric field. This observation is not necessarily surprising, since layer potential differences will induce a topological phase transition at the single-particle level, making the topmost valence band topologically trivial. However, our calculations show that FCI is suppressed well before the single-particle topological phase transition. In Fig.~\ref{fig:electric_field}(a), we show that the many-body gap closes at electric field $E=1.26$~meV/\AA{}. In Fig.~\ref{fig:electric_field}(b), we present the single particle band structure for $E=1.39$~meV/\AA{}, where an isolated flat Chern band can still be observed. The band width of the Chern band at $E=1.39$~meV/\AA{} is comparable to that at $E=0.0$~meV/\AA{}, but the FCI is already destroyed. In experiment, the ferromagnetism disappears at $E\sim 5$~meV/\AA{}~\cite{cai_signatures_2023}, implying FCI is destroyed at a smaller electric field. Our critical electric field $E=1.26$~meV/\AA{} is consistent with this observation.

It is well-known that completely quenched kinetic energy (i.e. vanishing band width) does not ensure the existence of FCI, and a number of proposals~\cite{PhysRevB.102.165148,PhysRevB.85.241308,PhysRevB.90.165139,PhysRevLett.127.246403,ledwith2022vortexability} are put forward to identify the conditions for FCI to emerge. Many of these proposals aim to design wave functions in a flat Chern band such that they closely resemble the wave functions of a Landau level. For example, since the Berry curvature $\Omega$ and quantum metric tensor $g$ is constant for Landau levels, the flatness of these two quantities~\cite{PhysRevB.102.165148,PhysRevB.85.241308,PhysRevB.90.165139} in the reciprocal space is heuristically viewed as a promising indicator for the emergence of FCI. In Fig.~\ref{fig:electric_field}(c), we present the distributions of $\Omega$ and $\mathrm{tr}(g)$ for $E=0.0$~meV/\AA{} and $E=1.39$~meV/\AA{}. A serious deterioration of the flatness of the Berry curvature and quantum metric tensor can be observed when the FCI is destroyed. Both $\Omega$ and $\mathrm{tr}(g)$ at $E=1.39$~meV/\AA{} are concentrated at the moire $K$ valley [Fig.~\ref{fig:electric_field}(c)], which explains the suppression of the FCI state.

\textit{Fractional Chern insulator at $\nu=-3/5$}.---In addition to the $\nu = -1/3$ state, the experimental observation also includes the $\nu = -3/5$ state~\cite{cai_signatures_2023}. However, in our calculations, we do not observe a clear many-body gap at $\nu = -3/5$ for the dielectric constant of $\epsilon = 15$. Nonetheless, when the dielectric constant is increased to $\epsilon = 8$, which is closer to the experimental value, we find evidence for the FCI~\cite{supplemental}. The characteristic interaction at $\epsilon = 8$ is larger than the energy gap between the two topmost bands, but the result can still be viewed as supporting evidences of the existence of FCI at $\nu = -3/5$.  We also performed ED calculations for $\nu= -1/5$, $-2/5$ and $-4/5$ and find no clear evidence of FCI for $\nu= -1/5$ and $-4/5$. For $\nu= -2/5$, we find evidence of a FCI phase with the many-body energy gap roughly half of that at $\nu=-3/5$.

In summary, we have presented a comprehensive theoretical study of the recent experimental observation of fractional Chern insulator at zero magnetic field in a twisted MoTe$_2$ bilayer. Density functional theory calculations reveal the existence of an isolated flat Chern band, which allows us to confirm the existence of a $\nu=-2/3$ fractional Chern insulator state using exact diagonalization. Phase diagrams are presented to guide future experimental study of fractional Chern insulator in this system, and the suppression of fraction Chern insulator by external electric field is studied. Our findings offer valuable insights into the nature and properties of fractional Chern insulators in moir\'e superlattices.

\begin{acknowledgments}
We thank Jie Wang, Jiaqi Cai and Heqiu Li for stimulating discussions. The exact diagonalization study is supported by DOE Award No. DE-SC0012509.  The density-functional theory calculation is supported by the Center on Programmable Quantum Materials, an Energy Frontier Research Center funded by DOE BES under award DE-SC0019443. YH acknowledges support from the European Research Council (ERC) under the European Union Horizon 2020 Research and Innovation Programme (Grant Agreement Nos. 804213-TMCS). This work was facilitated through the use of advanced computational, storage, and networking infrastructure provided by the Hyak supercomputer system and funded by the University of Washington Molecular Engineering Materials Center at the University of Washington (DMR-1719797).
\end{acknowledgments}

\textit{Note added.}---We recently became aware of an independent work on similar topics~\cite{fu_note}.

\bibliography{main}

\end{document}

\title{Supplemental Material for Fractional Chern Insulator in Twisted Bilayer MoTe$_2$}

\author{Chong Wang}
\affiliation{Department of Materials Science and Engineering, University of Washington, Seattle, WA 98195, USA}
\author{Xiao-Wei Zhang}
\affiliation{Department of Materials Science and Engineering, University of Washington, Seattle, WA 98195, USA}
\author{Xiaoyu Liu}
\affiliation{Department of Materials Science and Engineering, University of Washington, Seattle, WA 98195, USA}
\author{Yuchi He}
\affiliation{Rudolf Peierls Centre for Theoretical Physics, Clarendon Laboratory, Parks Road, Oxford OX1 3PU, United Kingdom}
\author{Xiaodong Xu}
\affiliation{Department of Physics, University of Washington, Seattle, WA 98195, USA}
\affiliation{Department of Materials Science and Engineering, University of Washington, Seattle, WA 98195, USA}
\author{Ying Ran}
\affiliation{Department of Physics, Boston College, Chestnut Hill, MA, 02467, USA}
\author{Ting Cao}
\email{tingcao@uw.edu}
\affiliation{Department of Materials Science and Engineering, University of Washington, Seattle, WA 98195, USA}
\author{Di Xiao}
\email{dixiao@uw.edu}
\affiliation{Department of Materials Science and Engineering, University of Washington, Seattle, WA 98195, USA}
\affiliation{Department of Physics, University of Washington, Seattle, WA 98195, USA}

\maketitle

\textit{Details of DFT calculations}.---\emph{Ab initio} DFT calculations including spin-orbit coupling are performed with the \textsc{siesta} package~\cite{soler2002siesta}.  Optimized norm-conserving Vanderbilt pseudopotentials~\cite{hamann2013optimized} and Perdew-Burke-Ernzerhof exchange-correlation functional~\cite{perdew1996generalized} are used. In the relaxations of the monolayer unit cell and the moir\'e superlattice, a double-zeta plus polarization basis is chosen and van der Waals corrections within the D2 formalism~\cite{grimme2006semiempirical} are used. The moir\'e superlattice is fully relaxed until the force on each atom is smaller than 0.01~eV/\AA{}.

\begin{figure}
\centering
\includegraphics[width=0.220\textwidth]{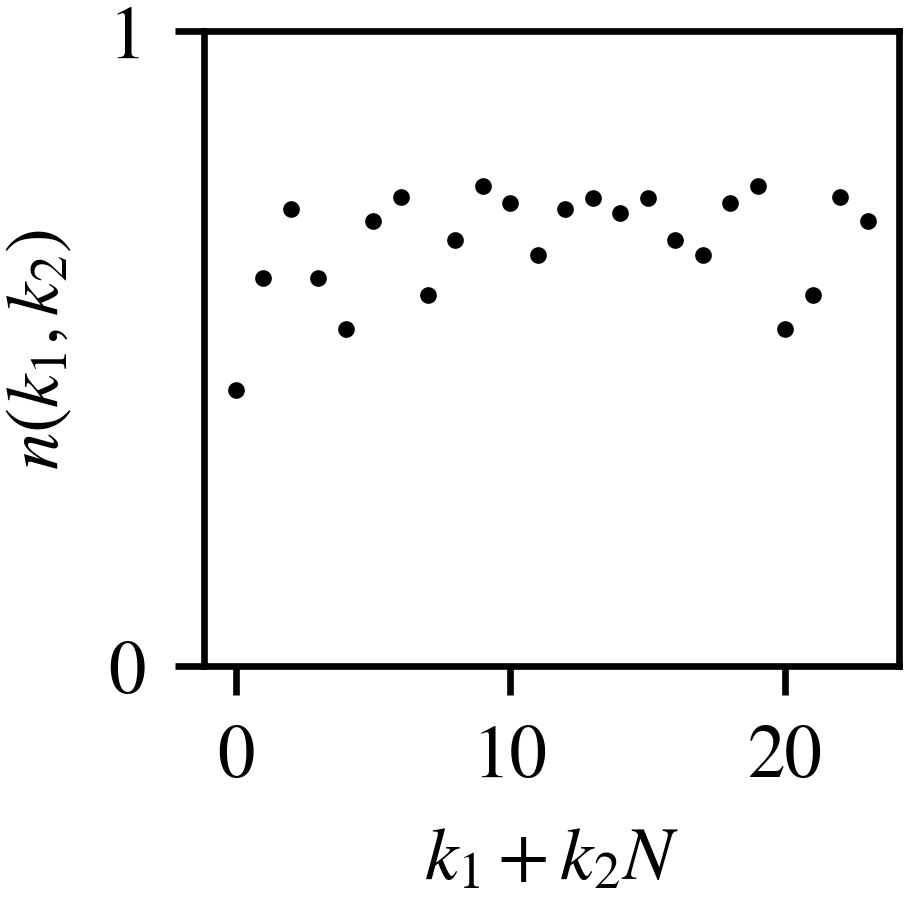}
\caption{The single particle occupation number $n(\bm{k})$ for the ground state at $\nu=-2/3$. $k_1$ and $k_2$ are indices of discretized momentums along $\bm{G}_1$ and $\bm{G}_2$. $N$ is the number of discretized momentums along $\bm{G}_1$. The calculation is carried out in a system with $4 \times 6$ unit cells.\label{fig:supp_occupation}
}
\end{figure}

\textit{Single particle occupation number at $\nu = -2/3$}.---Figure~\ref{fig:supp_occupation} shows the single particle occupation number $\langle n(\bm{k}) \rangle$ from the exact diagonalization calculation for the $\nu = -2/3$ state. The uniformity of $\langle n(\bm{k}) \rangle$ is a strong indicator favoring the fractional Chern insulator (FCI) state over charge density wave states, one of FCI's competing phases.

\begin{figure}
\centering
\includegraphics[width=0.439\textwidth]{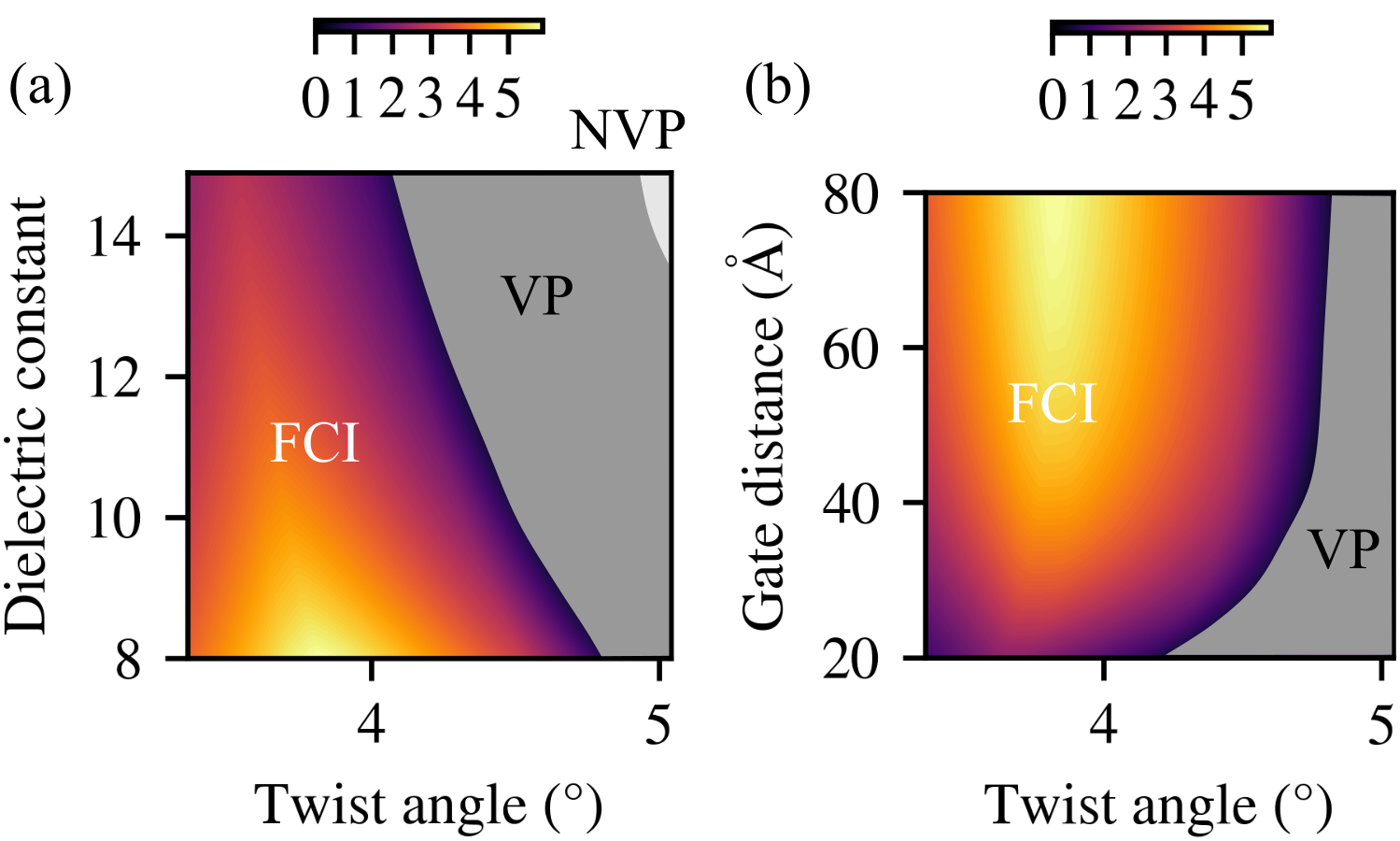}
\caption{Phase diagram as a function of twist angle and dielectric constant [(a)] or the distance between the sample and the gate [(b)]. $\epsilon$ is chosen as 8 for (b). Many-body gap is shown in both (a) and (b) by color in FCI phase (unit: meV). FCI is identified with $4 \times 6$ unit cells. Valley polarization is identified in a system with $3 \times 4$ unit cells.\label{fig:supp_phase_diagram}
}
\end{figure}

\textit{Phase diagram for smaller dielectric constant}.---In the main text, we only showed the phase diagram for the dielectric constant $\epsilon$ larger than 15 [Figure 3(a)] so that the characteristic interaction strength is smaller than the averaged energy gap between the two topmost bands.  In Fig.~\ref{fig:supp_phase_diagram}(a), we present the phase diagram for dielectric constant smaller than that chosen in the main text, with the caveat that the single band projection might not be accurate and treatment to include the band mixing is needed. The phase diagram as a function of twist angle and the distance between the sample and the gate is presented in Fig.~\ref{fig:supp_phase_diagram}(b) for $\epsilon=8$. There is no non-valley polarized phase in the parameter range.

\begin{figure}
\centering
\includegraphics[width=0.498\textwidth]{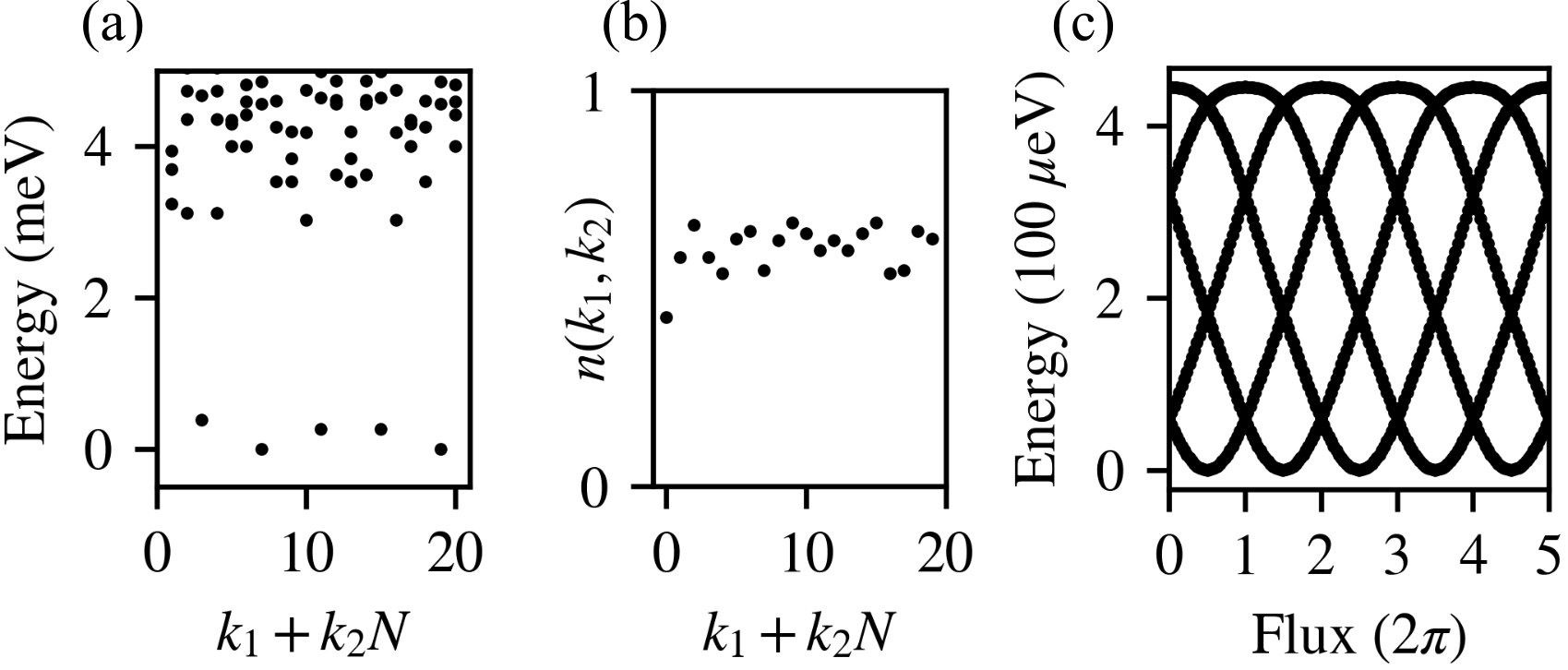}
\caption{Fractional Chern insulator phase at $\nu=-3/5$. The many body spectrum with the assumption of valley polarization shown in (a). The single particle occupation number is plotted in (b). (c) shows the evolution of ground states under flux insertion. During the flux insertion, many-body gap is maintained. The calculation is carried out in a system with $4 \times 5$ unit cells; the dielectric constant is chosen to be 8; the distance between gate and sample is chosen to be $d = 300$~\AA{}; the twist angle is 3.89$^\circ$.
\label{fig:supp_3_5}
}
\end{figure}

\textit{Exact diagonalization at $\nu = -3/ 5$}.---The many-body spectrum, single particle occupation number and the ground states' evolution under flux insertion are shown in Fig.~\ref{fig:supp_3_5} for $\nu = -3/5$, supporting the existence of the FCI phase.

\begin{figure}
\centering
\includegraphics[width=0.497\textwidth]{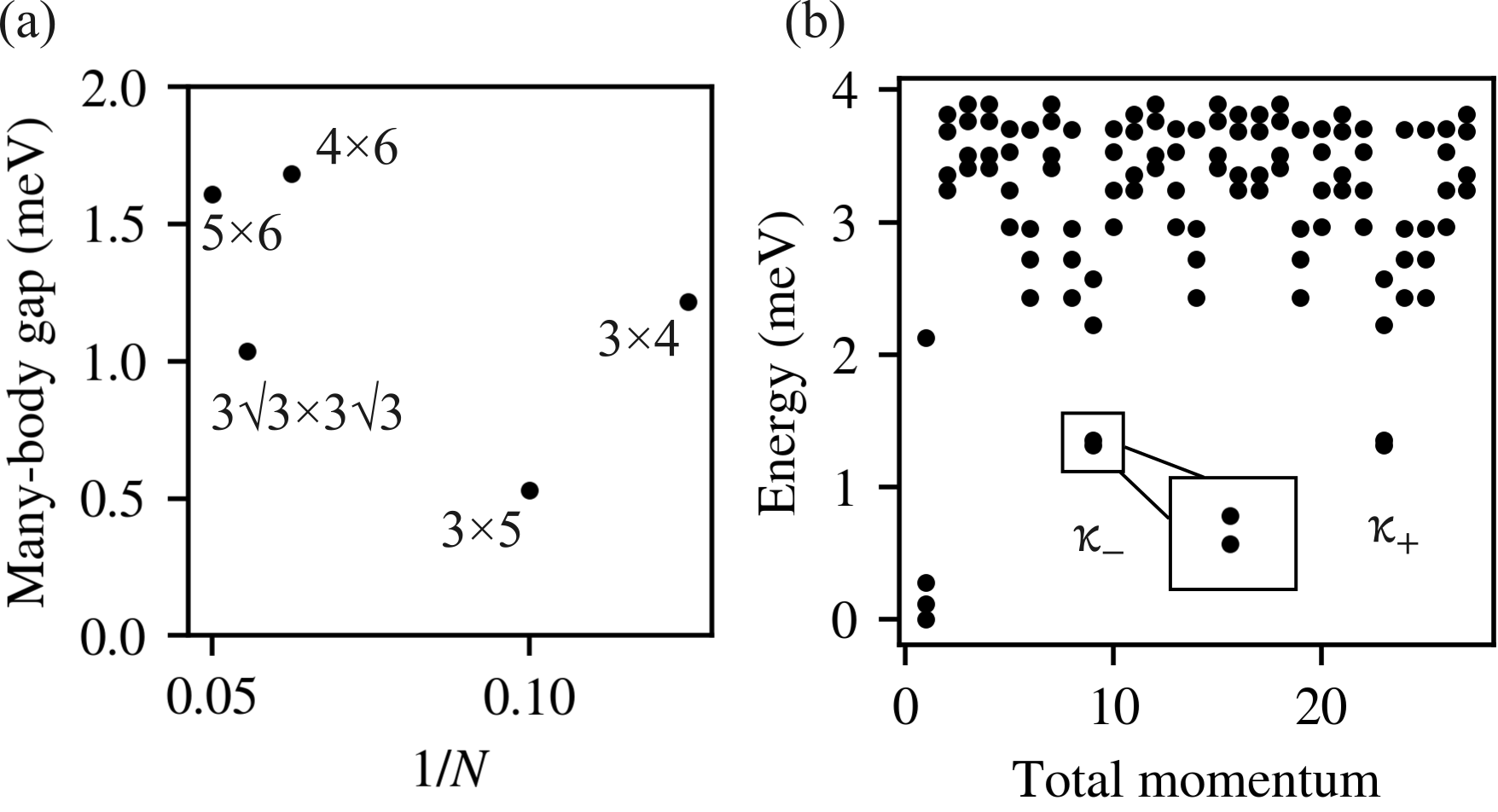}
\caption{(a) Many-body gap for various system sizes in exact diagonalization calculations for $\nu = -2/3$. The horizontal axis is inverse of the number of holes in the system. (b) The many-body spectrum of $\nu = -2/3$ on a system size of $3\sqrt{3} \times 3\sqrt{3}$. The computing parameters are the same as Fig.~2 in the main text.
\label{fig:supp_size}
}
\end{figure}

\textit{System size dependence of the exact diagonalization calculations}.---The many-body gap, defined as the energy difference between the fourth and third lowest eigenstates, is presented in Fig.~\ref{fig:supp_size}(a) for various system sizes. A smaller many-body gap is observed in the $3 \times 5$ system in comparison to the $3 \times 4$ system. The system is no longer in a FCI state for system size $3 \times 6$ for $\epsilon = 15$ (for $\epsilon = 8$, the features for FCI is still present). This can be attributed to the two-dimensional nature of FCI: many-body Chern number cannot be defined in a one-dimensional system. Systems with larger size and smaller aspect ratio all show large many-body gaps. For systems with a  $3\sqrt{3} \times 3\sqrt{3}$ size, $\sqrt{3}\times\sqrt{3}$ supercells is first constructed, and the supercells are then repeated in two lattice directions 3 times. These systems are compatible with three-fold rotational symmetric charge density wave (CDW) states and the many-body spectrum is presented in Fig.~\ref{fig:supp_size}(b). There seems to be a tendency for the system to form CDW states based on low-energy excitations. However, in our calculations, CDW states are never unambiguously stabilized.

\begin{figure}
\centering
\includegraphics[width=0.70\textwidth]{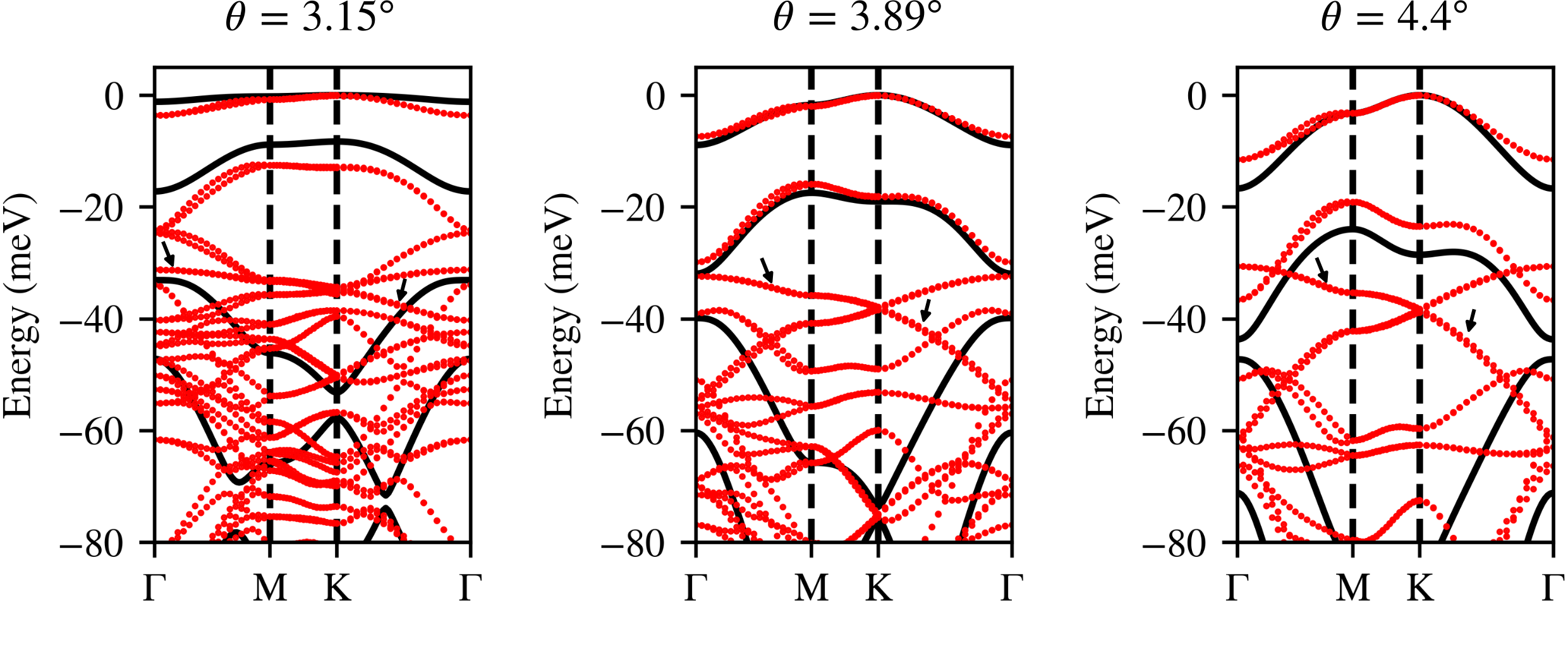}
\caption{Band structures for twisted bilayer MoTe$_2$ with various twist angles. The black lines are from continuum models with parameters in the main text. The red dots are from DFT calculations. The two topmost bands from the $\Gamma$ valley are labeled by black arrows.\label{fig:supp_bands}
}
\end{figure}

\textit{Band structures for different twist angles}.---In Fig.~\ref{fig:supp_bands}, we present band structures for twisted bilayer MoTe$_2$ for different twist angles. The continuum model (black lines) are fitted at twist angle $\theta = 3.89^\circ$. At twist angles $\theta = 3.15^\circ$ and $\theta = 4.4^\circ$, the bands from the continuum model deviates from DFT bands, even for the two topmost moir\'e bands. For the topmost band, the band width decreases with decreasing twist angle. However, the continuum model appears to provide an overestimation for the rate at which the band width changes with respect to the twist angle. The overestimation likely stems from the inability to accurately extrapolate the effects of lattice relaxation at $\theta = 3.89^\circ$ to other twist angles.

In the main text, our calculations choose $3.4^\circ < \theta < 5.0^\circ$. The lower bound is relatively close to $3.89^\circ$, a point at which our model demonstrates accuracy. Therefore, the claim of 3.5$^\circ$ should be relatively accurate. Conversely, the upper bound may be considered relatively less precise. We suspect that the FCI phase could remain stable even for twist angles slightly larger than those we have depicted in the phase diagram.

\begin{figure}
\centering
\includegraphics[width=0.250\textwidth]{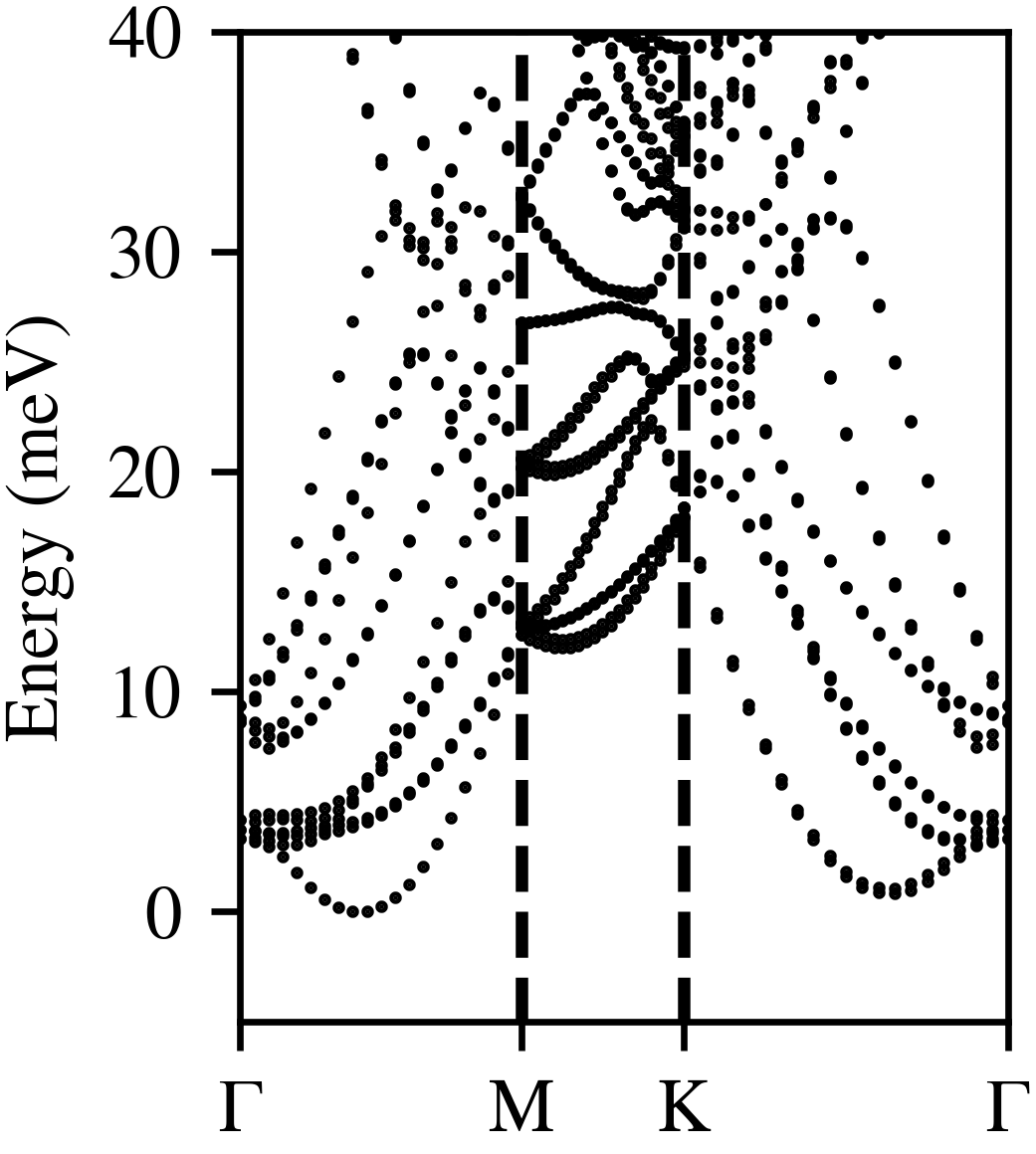}
\caption{Conduction band structures for twisted bilayer MoTe$_2$ at $\theta = 3.89^\circ$\label{fig:supp_bands_cond}
}
\end{figure}

\textit{Conduction bands of twisted bilayer MoTe$_2$}.---In Fig.~\ref{fig:supp_bands_cond}, we present the DFT conduction bands for twisted bilayer MoTe$_2$ at $\theta = 3.89^\circ$. The conduction bands are mainly from the $Q$ valley. Experimentally, non-topological insulators are observed in electron doping side~\cite{cai_signatures_2023}. The band structure might be useful to other studies for optical/excitonic physics of the system. The band gap is 0.9~eV, but DFT usually underestimates band gap. 

\begin{figure}
\centering
\includegraphics[width=0.90\textwidth]{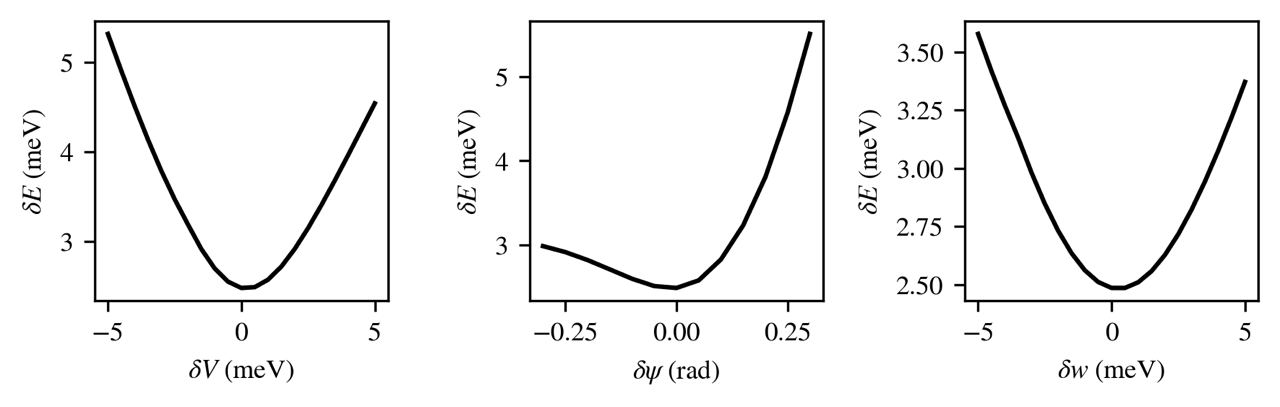}
\caption{
The loss function $\delta E$ as a function of the deviation of the three continuum model parameters.\label{fig:supp_parameters}
}
\end{figure}

\textit{Fitting of the continuum model}.---To obtain the continuum model parameters, we fit the band energies from the continuum to the DFT band energies on a $6 \times 6$ mesh in the Brillouin zone. To fit the band energy, we define the loss function as $\delta E =  \sqrt{\sum_{n,\bm k}[E^{\rm continuum}_n(\bm{k}) - E^{\rm DFT}_n(\bm{k})]^2 / N}$ and minimize the loss function with respect to the continuum model parameters. Here, $E^{\rm continuum/DFT}_n(\bm{k})$ is the energy from the continuum model/DFT calculations for band $n$ at $\bm{k}$; $N$ is the total number of data points. To test the sensitiveness of the fitting, we tune the parameters away from the local minimum and present the loss function in Fig.~\ref{fig:supp_parameters}. It can be observed that all the parameters are relatively sensitive to the fitting. We emphasize that with 72 data points and only 3 parameters, the fitting is well within a regime that avoids overfitting.

% In addition, we have calculated the eigenvalues of three-fold rotational symmetry at high symmetric points from DFT wave functions, and have deduced the Chern numbers from the eigenvalues (cf. Tab.~\ref{table:chern}). The Chern numbers from DFT calculations are consistent with the Chern numbers from continuum model. We have also checked that the real space distribution of the wave function is consistent between DFT calculations and continuum model calculations.

% \begin{table}
% \begin{tabular}{|c|c|c|c|c|} 
% \hline
% & $\Gamma$ & $K$ & $K'$ & Chern number (mod 3)\\ 
% \hline
% Topmost band & $\exp(i\pi)$ & $\exp(i\pi/3)$ & $\exp(i\pi/3)$ & $1$ \\
% \hline
% Second topmost band & $\exp(i\pi)$ & $\exp(-i\pi/3)$ & $\exp(-i\pi/3)$ & $-1$\\ 
% \hline
% \end{tabular}
% \caption{Three-fold rotational symmetry eigenvalues at high symmetric points in the moir\'e reciprocal space for twisted bilayer MoTe$_2$ at $\theta = 3.89^\circ$. The bands are from the $K$ valley. Bands from the $K'$ valley have opposite Chern numbers.}
% \label{table:chern}
% \end{table}

\clearpage

\bibliography{main}